\documentclass[apjl]{emulateapj}

\usepackage{graphicx}
\usepackage{apjfonts}
\slugcomment{Submitted to ApJ}
\shorttitle{The connection between Internetwork magnetic elements and supergranular flows}
\shortauthors{Orozco Su\'arez, Katsukawa, and Bellot Rubio}

\usepackage{color}

\begin{document}

\title{THE CONNECTION BETWEEN
INTERNETWORK MAGNETIC ELEMENTS AND
SUPERGRANULAR FLOWS}

\author{D.\ Orozco Su\'arez, Y.\ Katsukawa}

\email{dorozco@nao.ac.jp}

\affil{National Astronomical Observatory of Japan, 2-21-1 Osawa, Mitaka, Tokyo 181-8588, Japan}

\and

\author{L.~R.\ Bellot Rubio}

\affil{Instituto de Astrof\'{\i}sica de Andaluc\'{\i}a (CSIC), Apdo.\ de Correos 3004, 18080 Granada, Spain}

\begin{abstract}
The advection of internetwork magnetic elements by supergranular
convective flows is investigated using high spatial resolution, high
cadence, and high signal-to-noise ratio \ion{Na}{1}~D1 magnetograms
obtained with the \emph{Hinode} satellite. The observations show that
magnetic elements appear everywhere across the quiet Sun surface. We
calculate the proper motion of these magnetic elements with the aid of
a feature tracking algorithm. The results indicate that magnetic
elements appearing in the interior of supergranules tend to drift
toward the supergranular boundaries with a non-constant velocity. The
azimuthally averaged radial velocities of the magnetic elements and of
the supergranular flow, calculated from a local correlation tracking
technique applied to Dopplergrams, are very similar. This suggests
that, in the long term, surface magnetic elements are advected by
supergranular flows, although on short time scales their very chaotic
motions are driven mostly by granular flows and other processes.
\end{abstract}


  \section{Introduction}
  \label{sec1}

\begin{figure*}[!t]
\begin{center}
\epsscale{0.8}
\plotone{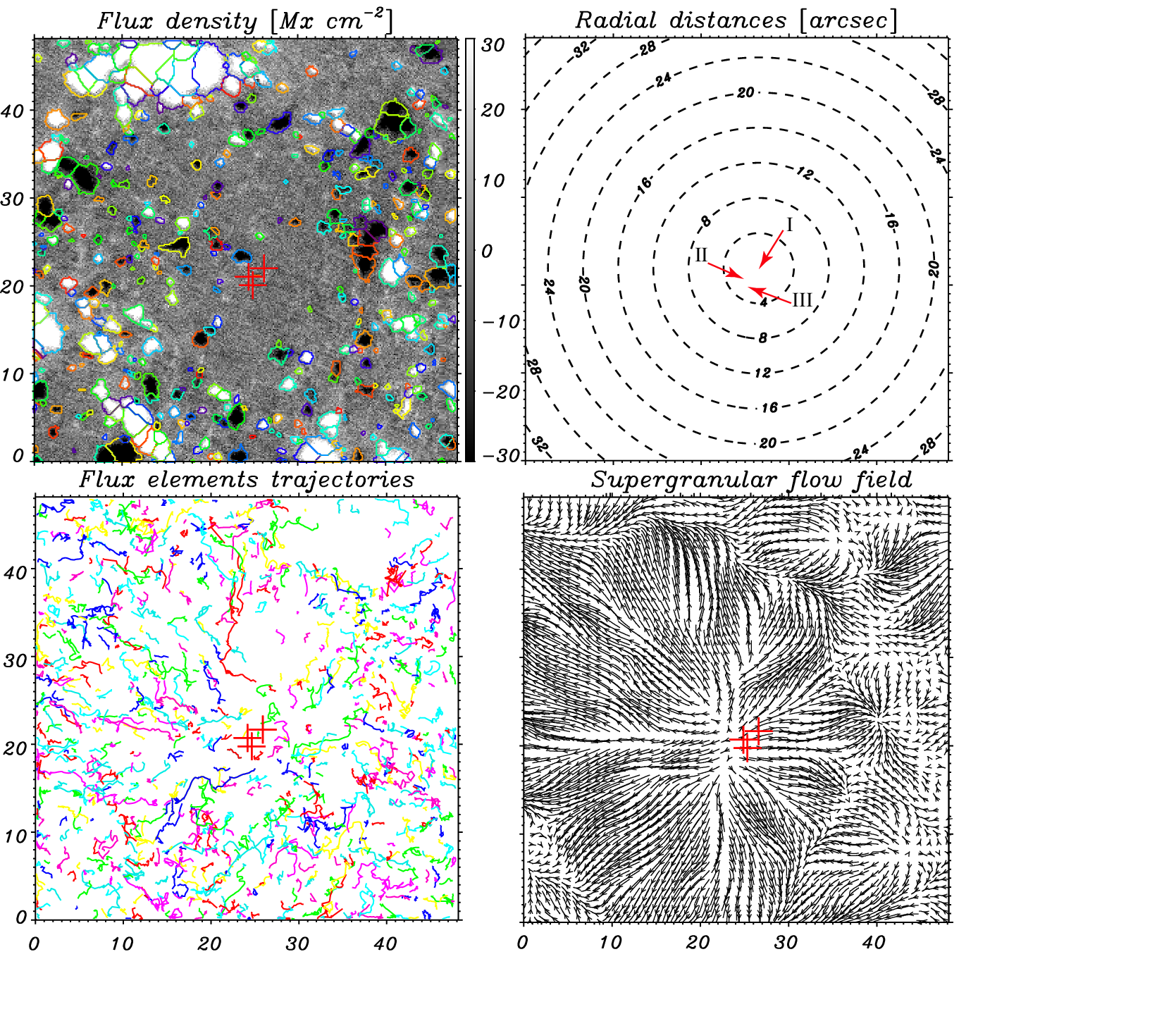}
\end{center}
\caption{\emph{Top-left}: image from the time sequence of high-resolution magnetograms. A mpeg animation is available in the electronic edition of the {\it Astrophysical Journal}. It shows the rapid flux appearance and disappearance of IMEs and their drift toward stronger flux concentrations. Colors outline magnetic features detected with the tracking algorithm. The magnetograms are saturated at $\pm$~30~G. \emph{Top-right}: Contours of constant distance computed from the supergranular center (I). \emph{Bottom-left}: Paths of some of the IMEs detected and tracked using YAFTA\_10 in 13 hours time series of magnetograms. They are rather lineal in the inner part and more random near the borders. Color code has no meaning.  \emph{Bottom-right}: Horizontal velocity field (stream-lines) derived from the motion of granules for a 8 hours window. Axis units are in arcsecond. Arrows in the top-right panel point to the spatial locations of centers I, II, and III (see text). The locations of the centers are represented by red crosses in the rest of the panels.}
\label{fig1}
\end{figure*}

Convection is probably the most efficient mechanism for energy transport in the solar
surface \citep{1990ARA&A..28..263S}. It works in the same way on all spatial scales but, due to the large stratification of the solar atmosphere, its properties change dramatically with height (see \citealt{2010LRSP....7....2R} and \citealt{2009LRSP....6....2N} for reviews). 
For instance, the average diameter of convective cells varies from the 1 Mm of the
granules observed at the surface to the 20-30 Mm of the supergranular cells at tens 
of Mm below the surface that have been inferred from observations, theory, and numerical
simulations (e.g. \citealt{2003ApJ...597.1200R}). The average lifetime also varies from 10 minutes to hours or days. There is an intermediate
scale -mesogranulation- whose existence and origin are still under
debate (see e.g., \citealt{2011ApJ...727L..30Y}).  First estimates of
the horizontal velocity component of the supergranular flow were obtained by
\cite{1964ApJ...140.1120S} and \cite{1962ApJ...135..474L}. They reported velocities of about 0.3~km~s$^{-1}$ with a peak of
0.5~km~s$^{-1}$. Recent investigations have concentrated on the
determination of the weaker vertical component of the supergranular
flow, which is of order 10~m~s$^{-1}$ only (see,
e.g., \citealt{2010ApJ...725L..47D}).

It has been known for long that there exists a close relationship
between supergranulation and the fields which make up the quiet Sun magnetic network \citep{1964ApJ...140.1120S}. We also know that most
magnetic field elements pervading the quiet Sun surface are very weak
\citep{Orozco1,Orozco2}. These fields can only be advected and transported
across the surface by three different flow patterns: granulation,
mesogranulation, and supergranulation. Indeed, since the discovery of
magnetic fields in the interior of supergranular (network) cells by
\cite{1971IAUS...43...51L,1975BAAS....7..346L} it is known that 
internetwork magnetic elements (hereafter IMEs) drift toward strong
magnetic concentrations with a mean horizontal speed\footnote{Defined
as $\sqrt{v_x^2+v_y^2}$, with $v_x$ and $v_y$ the two orthogonal
components.} of 0.5~km~s$^{-1}$. The proper motion of IMEs can be separated in two main components\footnote{Here we ignore the drifts of IMEs due to mesogranular flows and the motion of the footpoints of
emerging loops: they show a ballistic velocity of 3~km~s$^{-1}$ during
the emergence phase, i.e., when they move from the granule toward an adjacent 
intergranular lane \citep{2009ApJ...700.1391M,2011A&A...531L...9M}.}: one of random nature resulting from the continual buffeting of the fields by granular convection \citep{2011A&A...531L...9M} and a systematic motion that transports the
fields toward strong flux concentrations delineating the supergranular borders. 
The horizontal velocity of IMEs shows an rms fluctuation of about
1.6~km~s$^{-1}$ \citep{2008ApJ...684.1469D}. The relative contribution
of the random and systematic velocities to this enormous fluctuation
is still under debate. The values reported in the literature for the
net velocity component range from 0.2~km~s$^{-1}$
\citep{2008ApJ...684.1469D} to 0.35~km~s$^{-1}$
\citep{1987SoPh..110..101Z}. The net velocity may depend on the
location of the IMEs within the supergranular cell and even on the
polarity of the IMEs and the nearby network
fields. \cite{1998A&A...335..341Z} investigated the spatial dependence
of the velocity of IMEs and found a constant radial velocity (measured
from the center of supergranular cells outward) of about
0.4~km~s$^{-1}$ using low resolution (1\farcs5) and low cadence (7
minutes) magnetograms. Recently, \cite{2008ApJ...684.1469D} have argued
that the net velocity of the IMEs may be larger in the proximity
of network fields of opposite polarity, suggesting the existence of
some kind of magnetic ``attraction'' between IMEs and the
network.

Unambiguous observational proofs that IMEs are advected by convective
flows are still missing. In this Letter we use high spatial resolution
magnetograms acquired with the \emph{Hinode} satellite
\citep{2007SoPh..243....3K} to investigate the relationship between
the net velocity of IMEs and the horizontal supergranular flow.

\section{Observations, magnetic features tracking, and local correlation tracking}
\label{sec2}

The data used here belong to the Hinode Operation Plan 151, entitled
``Flux replacement in the photospheric network and internetwork''. 
They consist of shutterless Stokes I and V filtergrams taken on 
2010 April 20 at disk center\footnote{The heliocentric angle
varied $\sim$~0.01 units during the observing time.} with the \emph{Hinode}
Narrowband Filter Imager (NFI; \citealt{2008SoPh..249..167T}) in the
two wings of \ion{Na}{1}~D1 589.6~nm, $\pm 16$~pm apart from 
the line center. The total integration time for each wavelength and 
polarization state was 6.4~s. It took about 38~seconds to complete a
93\arcsec$\times$82\arcsec\/ map in the shutterless mode. The final
cadence was set to 80 seconds. The diffraction-limited resolution of
the NFI filtergrams at 589.6~nm is 0\farcs24 ($\sim$~175~Km) and the pixel size is
0\farcs16, thus the data are slightly over-sampled although the Lyot
filtergraph worsen the spatial resolution to about 0\farcs3 ($\sim$~215~Km).  We used
13 hours of observations, about half of the mean lifetime of supergranules
(16--23 hr; e.g., \citealt{2004ApJ...616.1242D}).

\begin{figure*}[!t]
  \centering
\epsscale{0.45}
\plotone{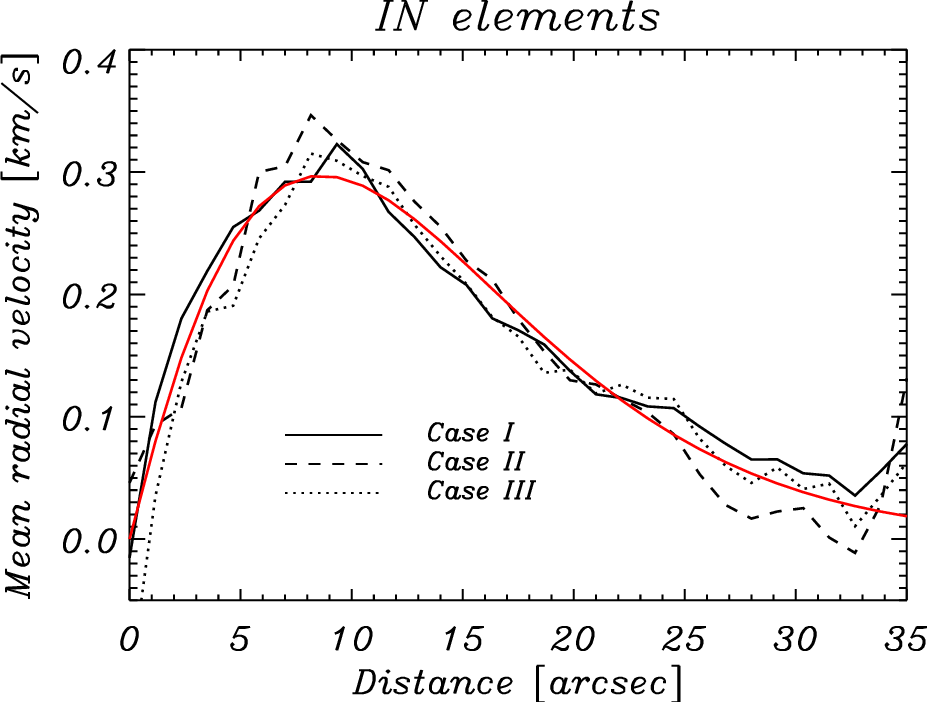}  
\plotone{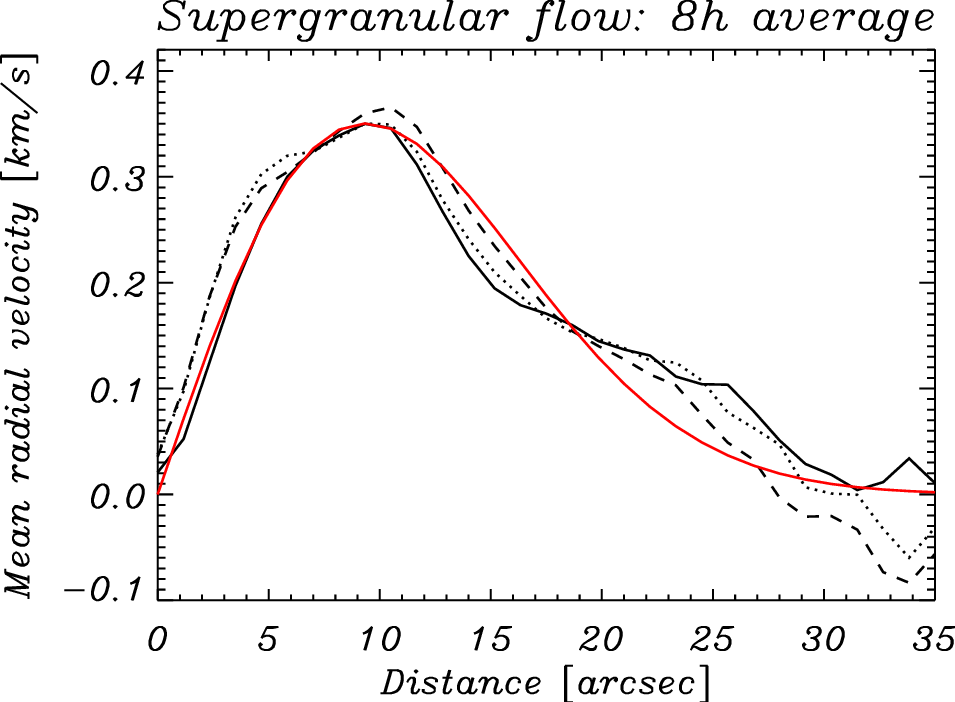}
\plotone{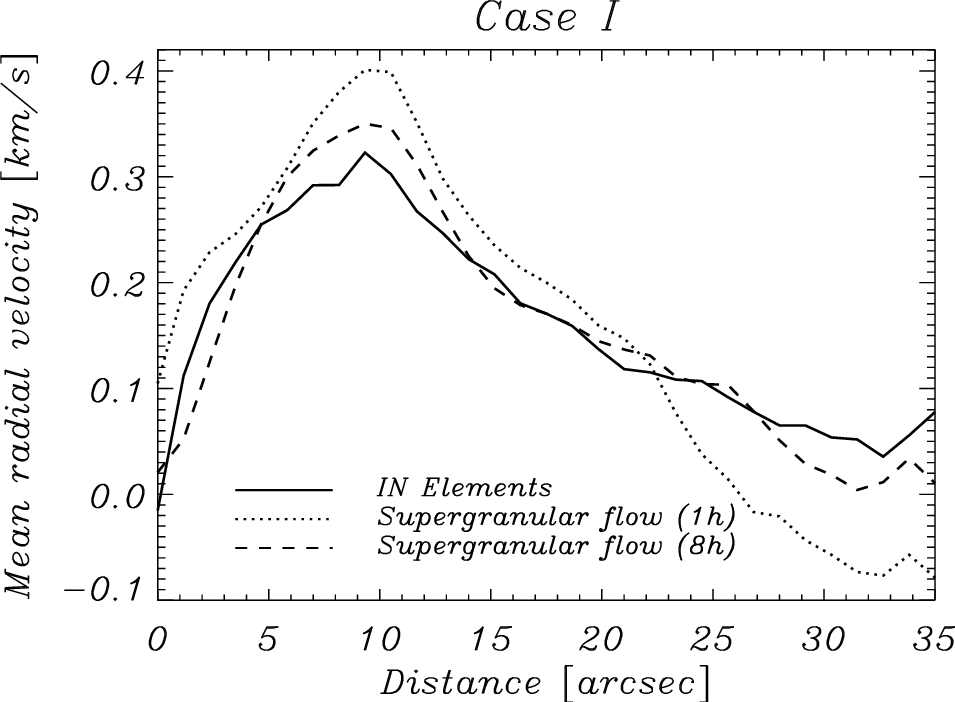}
\plotone{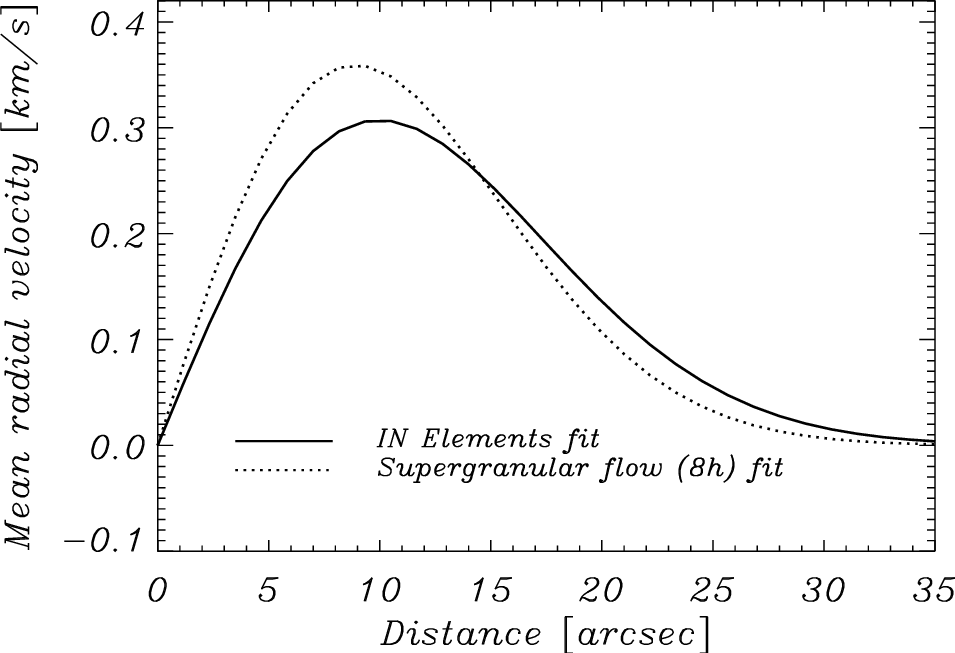}
\caption{Averaged radial (outward) velocities against radial distance from the center of the supergranule. Top  panels represent the radial velocities calculated from the tracking of IMEs (left) and from FLCT (right) using three different centers. The red curves represent fits to a simple supergranular kinematic model for center I (see the text). The bottom left panel shows a comparison of the radial velocity calculated from the IMEs (solid line) with the supergranular flow velocity (dotted and dashed lines, 1 and 8 hr averaged, respectively) using center I.  The fits are put together in the bottom right panel for comparison.}
\label{fig2}
\end{figure*}

The data were calibrated as follows: first we obtained magnetograms and
Dopplergrams as $M=0.5\times (V_b/I_b-V_r/I_r)$ and
$D=(I_r-I_b)/(I_r+I_b)$, with $I$ and $V$ representing the intensity and
the circular polarization signal, and $r$ and $b$ the red and blue
wings of the line. We then removed the (residual) effects of the
transmission profile shift resulting from the satellite orbital
variation. Finally, we applied a subsonic filter (see e.g.,
\citealt{1997ApJ...480..406H}) to the data to remove p-mode oscillations
with periods of 5 minutes.  The magnetogram signal $M$ was converted into
flux density multiplying it by the calibration constant $C=9160$ derived
from the weak field approximation \citep{milan}. No additional corrections were
applied to the Dopplergrams because they are used only for local
correlation tracking. The final rms noise of the magnetograms is
6.7G, allowing the detection of very weak IMEs. Figure~\ref{fig1}
(top left panel) depicts a single magnetogram of
48\arcsec$\times$48\arcsec. The stronger circular polarization
signals are concentrated near the borders of the image. These fields
correspond to the network that, in general, outlines supergranular
boundaries. Thus, the image roughly encompasses one supergranular
cell. In the interior, there are small polarization patches that show
opposite polarities. These signals correspond to IMEs.

For detecting and tracking IMEs we used the YAFTA\_10 (Yet Another
Feature Tracking Algorithm) software\footnote{YAFTA\_10, written in IDL, 
can be downloaded from the author's Web site at 
http://solarmuri.ssl.berkeley.edu/$\sim$welsch/public/software/YAFTA/} 
\citep{2007ApJ...666..576D}. We ignored all magnetic elements 
with flux densities below 10~G and smaller than 16~pixel$^2$. Colors in
the top-left panel of Figure~\ref{fig1} represent the magnetic
elements detected by YAFTA\_10 in a single frame. The animation shows
the continuous appearance and disappearance of IMEs of both polarities
in the quiet solar surface and how the tracking software detects and
follows them. Interestingly, the visual impression is one of IMEs
moving radially toward the network. Indeed, as can be seen in the
bottom left panel of Figure~\ref{fig1}, the trajectories of the
magnetic elements located in the interior of the supergranule are
rather linear and radially aligned with respect to the center of the field of view
(FOV). The paths located in the network look more random and show little
linear trends. Using the tracking results, we derived the horizontal
velocity of the magnetic elements. To compute the velocities, we
neglected IMEs that did not last longer than three frames, leaving us
with a total of 17\,322 IMEs, of which only 7\,844 are
within a 20\arcsec\/ circle from the center of the FOV.

To calculate the horizontal velocity of the flows we applied a
Fourier Local Correlation Tracking method (FLCT;
\citealt{2008ASPC..383..373F}) to consecutive Dopplergrams. The
algorithm cross-correlates two images by comparing small
subfields. The output is the local displacements of one image with
respect to the other. The subfields were defined by a Gaussian window
of full-width at half-maximum 10~px (about 1\farcs5), roughly
corresponding to granules that are the best tracers of large-scale
velocity flows
\citep{2001A&A...377L..14R}. This process is done for the full time
series. The two components of the horizontal flow velocity are then calculated from running averaged FLCT flow-field maps. Since the horizontal flow pattern might slightly change depending on the width of the window and the total number of images used for the averaging, we have calculated the horizontal velocity for two different running averaged flow-field maps corresponding to about 1 and 8 hours duration. The bottom right panel of Fig.~\ref{fig1}
displays the stream-lines (flow map) calculated from the 8 hours averaged FLCT maps. Clearly, the flow goes from the center of the FOV outward.

Finally, we obtain the radial and transverse components of the
horizontal velocities with respect to the supergranular center, for
both IMEs and the supergranular flow. The radial component is defined
to be positive outward. We locate the center of the supergranule at 
the position of minimum radial velocity calculated from the magnetic
element tracking, at the position of minimum horizontal velocity of
the 8 hours averaged flow map, and the same for the 1 hour averaged flow map:
C$_{\rm I} \simeq $~(26\farcs6~,~22\farcs1), C$_{\rm II}
\simeq$~(24\farcs8~,~21\farcs1), and C$_{\rm III} \simeq
$~(25\farcs3~,~20\farcs2), respectively. Next we calculate azimuthally
averaged values of the radial velocity component as a function of
distance from the center of the supergranular cell (see the top right
panel of Figure~\ref{fig1}).

\section{Results}\label{sec3}

Figure~\ref{fig2} shows the averaged radial velocity versus radial
distance from the center of the supergranule calculated from the
tracking of IMEs and from the FLCT applied to the Dopplergrams.  These
values are averages over an extremely large range of individual radial
velocities (the typical rms fluctuation in individual bins is 0.7~km~s$^{-1}$).  The average radial component of the IME velocity shows a clear
dependence with distance to the supergranular cell center (see top
left panel). The magnetic elements near the center of the supergranule are
almost at rest. Their radial velocity increases outwards
until it reaches a maximum of 0.3~km~s~$^{-1}$ at 9\arcsec\/ from the
supergranular center. Then, it decreases monotonically. The mean
radial velocity, calculated by averaging the mean radial velocities over radial distance, is about 0.15~km~s~$^{-1}$, smaller than the 0.4~km~s~$^{-1}$ estimated\footnote{Note that \cite{1998A&A...335..341Z} directly averaged the radial velocities of individual IMEs to give a mean value while  \cite{2008ApJ...684.1469D} use an autocorrelation analysis.} by \cite{1998A&A...335..341Z} and the 0.2~km~s~$^{-1}$ of \cite{2008ApJ...684.1469D}. Remarkably, the average velocity is always positive, meaning that IMEs
tend to move systematically toward the network. On short time scales,
IMEs can move in any direction with much larger speeds, as revealed by
the very large spread of the individual velocities, which can be both
positive and negative.  

The top right panel of Fig.~\ref{fig2} shows the radial velocity
associated with the supergranular convection as determined from FLCT
with an 8 hour average. The velocity profile shape and its
variation are remarkably similar to those calculated from the tracking
of IMEs. The rms fluctuation in individual bins is 0.3~km~s$^{-1}$ on average. To stress the similarities between the curves we represent both
radial velocity profiles in the bottom left panel of the same figure. The radial velocity
determined from FLCT is slightly steeper although the exact variation
depends on the adopted time window. Indeed, the velocity profile
becomes gentler as the average window increases from 1 to 8 hours
(dotted and dashed lines, respectively). Overall, Fig.~\ref{fig2}
suggests that IMEs are continuously advected by large-scale convective
flows, although on short time scales the velocity is dominated by 
granular motions or other effects (as evidenced by the large rms fluctuations). 

To make the radial velocity profiles available to the community we have fitted
them using the supergranular convection
kinematic model of \cite{1989ApJ...345.1060S} (see also
\citealt{2001ApJ...561..427S}). In their model the radial (outward)
velocity is given by $v(r) = V(r/R)e^{-(r/R)^2}$ with $V$ representing
the amplitude of the radial velocity, $R$ the radius of the
supergranular cell, and $r$ the distance from the center of the
supergranule. This model satisfactorily represents the velocity
profiles with $V = 0.71\pm0.17$~km~s$^{-1}$ and $R=14.2\pm2.4$~Mm for the IMEs and
$V = 0.84\pm0.14$~km~s$^{-1}$ and $R=12.6\pm2.6$~Mm for the supergranular radial
flow field (see the red curves in Fig.~\ref{fig2}). To highlight the similarities and differences in the two fits we represent them in the bottom right panel of Fig.~\ref{fig2}. Thus, according
to this model, the radius of the supergranular cell is about 13~Mm.



\section{Conclusions}

We have shown that there exists a close correspondence between the
average radial velocity of IMEs and the horizontal velocity of
supergranular convective flows. Such a relation was often taken for
granted by the solar community because weak photospheric  fields should
be advected by surface flows, although there was no clear
observational proof of it. Our results show that the dynamic properties of
quiet Sun IMEs depend on their location within supergranular
cells. On average, IMEs first 
accelerate outwards and then decelerate. This behavior may affect the
distribution of magnetic flux within supergranular cells and also
suggests that magnetic diffusion may be more effective in certain
areas of the solar surface.

\acknowledgments
The authors thank the Japan Society for the Promotion of Science (JSPS) for the financial support through the postdoctoral fellowship program for foreign researchers. This work has been partly funded by the Spanish MICINN through projects AYA2011-29833-C06-04 and PCI2006-A7-0624, and by Junta de Andaluc\'{\i}a through project P07-TEP-2687, including a percentage from European FEDER funds. \emph{Hinode} is a Japanese mission developed and launched by ISAS/JAXA, with NAOJ as domestic partner and NASA and STFC (UK) as international partners. It is operated by these agencies in co-operation with ESA and NSC (Norway). Use of NASA's Astrophysical Data System is gratefully acknowledged.

\end{document}